\documentclass[twocolumn,showpacs,preprintnumbers,amsmath,amssymb,prl,floatfix]{revtex4}
\usepackage{graphicx,tabularx}
\usepackage{amssymb}
\usepackage{dcolumn}
\usepackage[mathcal]{euscript}
\begin{document}
\title{Work functions of self-assembled monolayers on metal surfaces}
\author{Paul C. Rusu}
\affiliation{Computational Materials Science, Faculty of Science and Technology and MESA+ Institute
for Nanotechnology, University of Twente, P.O. Box 217, 7500 AE Enschede, The Netherlands}
\author{Geert Brocks}
\affiliation{Computational Materials Science, Faculty of Science and Technology and MESA+ Institute
for Nanotechnology, University of Twente, P.O. Box 217, 7500 AE Enschede, The Netherlands}

\date{\today}

\begin{abstract}
Using first-principles calculations we show that the work function of noble metals can be decreased
or increased by up to 2 eV upon the adsorption of self-assembled monolayers of organic molecules.
We identify the contributions to these changes for several (fluorinated) thiolate molecules
adsorbed on Ag(111), Au(111) and Pt(111) surfaces. The work function of the clean metal surfaces
increases in this order, but adsorption of the monolayers reverses the order completely. Bonds
between the thiolate molecules and the metal surfaces generate an interface dipole, whose size is a
function of the metal, but it is relatively independent of the molecules. The molecular and bond
dipoles can then be added to determine the overall work function.
\end{abstract}


\maketitle
Recent advances in molecular electronics, where organic molecules constitute active materials in
electronic devices, have created a large interest in metal organic interfaces \cite{Kahn:jpsb03}.
Transport of charge carriers across the interfaces between metal electrodes and the organic
material often determines the performance of a device \cite{Mihailetchi:apl04}. Organic
semiconductors differ from inorganic ones as they are composed of molecules and intermolecular
forces are relatively weak. In a bulk material this increases the importance of electron-phonon and
electron-electron interactions \cite{Brocks:prl04}. At a metal organic interface the energy barrier
for charge carrier injection into the organic material is often determined by the formation of an
interface dipole localized at the first molecular layer. The interface dipole can be extracted by
monitoring the change in the metal surface work function after deposition of an organic layer
\cite{Kahn:jpsb03,Tal:apl04}.

Atoms and molecules that are physisorbed on a metal surface usually decrease the work function, as
the Pauli repulsion between the molecular and surface electrons decreases the surface dipole
\cite{Bagus:prl02,Dasilva:prl03}. Chemisorption can give an increase or a decrease of the work
function, and can even lead to counterintuitive results \cite{Michaelides:prl03,Leung:prb03}.
Self-assembled monolayers (SAMs) are exemplary systems to study the effect of chemisorbed organic
molecules upon metal work functions \cite{Campbell:prb96}. More specifically, alkyl thiolate
($\mathrm{C}_n\mathrm{H}_{2n+1}\mathrm{S}$) SAMs on the gold (111) surface are among the most
extensively studied systems
\cite{Schreiber:pss00,Alloway:jpcb03,Deboer:am05,Derenzi:prl05,Konopka:prl05}. The sulphur atoms of
the thiolate molecules form stable bonds to the gold surface and their alkyl tails are close
packed, which results in a well ordered monolayer. SAMs with similar structures are formed by alkyl
thiolates on a range of other (noble) metal surfaces \cite{Konopka:prl05,Schreiber:pss00,Lee:ss02}.

Often the change in work function upon adsorption of a SAM is interpreted mainly in terms of the
dipole moments of the individual thiolate molecules, whereas only a minor role is attributed to the
change induced by chemisorption \cite{Campbell:prb96,Alloway:jpcb03,Deboer:am05}. This assumption
turns out to be reasonable for adsorption of methyl thiolate (CH$_3$S) on Au(111)
\cite{Derenzi:prl05}, but for CH$_3$S on Cu(111) it is not \cite{Konopka:prl05}. In this paper we
apply first-principles calculations to study the interface dipoles and the work function change
induced by adsorption of thiolate SAMs.

In particular, we analyze the contributions of chemisorption and of the molecular dipoles to
uncover the effects of charge reordering at the interface. The chemical bonds between the thiolate
molecules and the metal surfaces generate an interface dipole. We find that this dipole strongly
depends upon the metal, but it is nearly independent of the electronegativity of the molecules. The
size and direction of the interface dipole are such that it overcompensates for the difference
between the clean metal work functions. This results in the SAM adsorbed on the highest work
function metal having the lowest work function and vice versa. Modifying the molecular tails allows
one to vary the absolute size of the work function over a range of more than 2 eV.

Since alkyl thiolate molecules form SAMs with a similar structure on (111) surfaces of several
noble metals, they are ideal model systems for studying metal organic interfaces. By varying the
relative electronegativity of surface and molecules one can induce electron transfer and create an
interface dipole, without completely rearranging the interface structure. The electronegativity of
a metal substrate is given by its work function. We consider the (111) surfaces of three metals
that have a substantially different work function, but the same crystal structure and a similar
lattice parameter: Ag (4.5 eV, 2.89\AA\ ), Au (5.3 eV, 2.88\AA\ ) and Pt (6.1 eV, 2.77\AA\ ).

One would also like to vary the molecule's electronegativity without changing the structure of the
SAM. This can be achieved by fluorinating the alkyl tails of thiolate molecules, which increases
their electronegativity \cite{Schreiber:pss00}. However, fluorinating the alkyl tails also reverses
the polarity of the thiolate molecules and one has to separate this electrostatic effect from the
charge reordering caused by chemisorption. In this paper we study the short chain thiolates
$\mathrm{C}\mathrm{H}_{3}\mathrm{S}$, $\mathrm{C}_2\mathrm{H}_{5}\mathrm{S}$,
$\mathrm{C}\mathrm{F}_{3}\mathrm{S}$, and
$\mathrm{C}\mathrm{F}_{3}\mathrm{C}\mathrm{H}_{2}\mathrm{S}$.

Density functional theory (DFT) calculations are carried out using the projector augmented wave
(PAW) method \cite{Bloechl:prb94,Kresse:prb99}, a plane wave basis set and the PW91 generalized
gradient approximation (GGA) functional, as implemented in the VASP program
\cite{Kresse:prb93,Kresse:prb96}. We use supercells containing a slab of at least five layers of
metal atoms with a SAM adsorbed on one side of the slab and a vacuum region of $\sim 12$ \AA. The
Brillouin zone of the $(\sqrt{3} \times \sqrt{3})\mathrm{R30^o}$ surface unit cell is sampled by a
$11 \times 11$ $\mathbf{k}$-point grid. The plane wave kinetic energy cutoff is 450 eV. To avoid
interactions between periodic images of the slab we apply a dipole correction
\cite{Neugebauer:prb92}. The geometry of the SAM is optimized, as well as the positions of the top
two layers of metal atoms. The atoms in the remaining metal layers are fixed at their bulk
positions. The optimized bulk lattice parameters are 2.93, 2.94 and 2.79 \AA\ for Ag, Au and Pt,
respectively.

The work function is given by $W=V(\infty)-E_F$, where $V(\infty)$ is the asymptotic electrostatic
potential in vacuum, and $E_F$ is the Fermi energy of the bulk metal. $V(\infty)$ is extracted from
the plane averaged potential $\overline{V}(z)=A^{-1}\int \!\!\! \int_A V(x,y,z)dxdy$, with $A$ the
area of the surface unit cell. In practice, $\overline{V}(z)$ reaches an asymptotic value within a
distance of 5 \AA\ from the surface. Accurate values of the Fermi energy are obtained following the
procedure outlined in Ref. \onlinecite{Fall:jp99}. By varying the computational parameters
discussed above we estimate that the work functions are converged to within 0.05 eV. Typically DFT
calculations give work functions that are within $\sim 0.1$-$0.2$ eV of the experimental values,
although occasionally somewhat larger deviations are found.

The $(\sqrt{3} \times \sqrt{3})\mathrm{R30^o}$ structure of CH$_3$S on Au(111) has been studied in
several first-principles calculations
\cite{Derenzi:prl05,Konopka:prl05,Yourdshahyan:prb01,Vargas:jpcb01,Hayashi:jcp01}. We find
basically the same optimized geometry as obtained in those calculations. Several structures exist
that have a slightly different geometry, but are very close in energy, such as a $c(4 \times 2)$
superstructure \cite{Vargas:jpcb01}. We find that the work functions of these structures are within
0.1 eV of that of the simpler structure, so we will not discuss these superstructures here.

The $(\sqrt{3} \times \sqrt{3})\mathrm{R30^o}$ structure is also a good starting point for studying
other systems. Thiolates with longer alkyl tails on Au(111) adopt this structure, as does CH$_3$S
on Pt(111), as well as alkyl thiolates on Au(111) whose end groups are fluorinated
\cite{Schreiber:pss00,Lee:ss02}. Thiolates with long alkyl tails on Ag(111) form a somewhat denser
packing, whereas long fluorinated alkyl thiolates form a somewhat less dense packing
\cite{Schreiber:pss00}. To analyze the work function we use optimized $(\sqrt{3} \times
\sqrt{3})\mathrm{R30^o}$ structures for all our SAMs. We find that varying the packing density only
introduces a scaling factor to the work function change \cite{Derenzi:prl05}.

\begin{table}[!tbp]
\caption{Calculated work functions $W$ (eV) of clean (111) surfaces and of surfaces covered by SAMs
in a $(\sqrt{3} \times \sqrt{3})\mathrm{R30^o}$ structure.\label{table:wf}}
\begin{ruledtabular}
\begin{tabular}{cccccc}
& clean & CH$_3$S & C$_2$H$_5$S & CF$_3$S & CF$_3$CH$_2$S \\ \hline
Ag & 4.50 & 3.95 & 4.13 & 6.14 & 6.30 \\
Au & 5.25 & 3.81 & 3.93 & 5.97 & 6.27 \\
Pt & 5.84(6.14\footnote{LDA value}) & 3.45 & 3.47 & 5.68 & 5.87 \\
\end{tabular}
\end{ruledtabular}
\end{table}

Table~\ref{table:wf} lists the calculated work functions. The work functions of the clean Au and Ag
surfaces agree with the experimental values \cite{Hansson:prb78,Monreal:jp03}, but that of Pt is
$\sim 0.3$ eV too low \cite{Derry:prb89}. The latter can be attributed to the GGA functional. Using
the local density approximation (LDA) the calculated work function of Pt(111) is 6.14 eV, which
agrees with experiment. In other cases the difference between the work functions calculated with
GGA and LDA functionals is much smaller. For instance, the GGA and LDA work functions of the SAMs
on Pt are within 0.02 eV of one another. We will use the GGA values throughout this paper. The
trend in the work functions of the SAM covered surfaces agrees well with experimental observations
\cite{Campbell:prb96,Alloway:jpcb03,Deboer:am05}. The experimental work function shifts with
respect to the clean surface are sometimes somewhat smaller than the calculated ones
\cite{fn:SAMexp}.

The first observation one can make by comparing the numbers in Table~\ref{table:wf} within columns
is that on SAM covered surfaces the work function \textit{decreases} in the order Ag, Au, Pt. This
is striking, since the work function of the clean metal surfaces clearly \textit{increases} in this
order. Secondly, comparing the numbers within rows one finds that the work functions of the
fluorinated alkyl thiolate covered surfaces are 2-2.5 eV higher than of the non-fluorinated ones.
We will argue that the first observation can be ascribed to the interface dipole formed upon
chemisorption. This interface dipole is independent of the molecular tails. The second observation
will be interpreted in terms of the individual molecular dipoles.

\begin{figure}[!tbp]
\includegraphics[width=6cm]{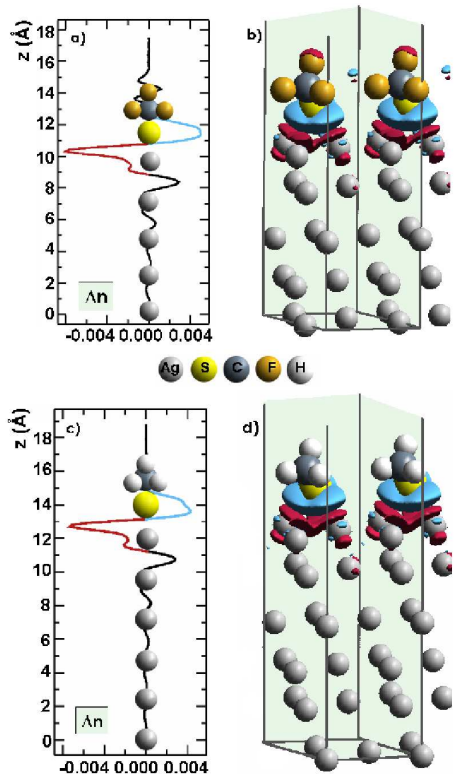}
\caption{ (Color online) Difference electron density $\Delta n = n_\mathrm{tot} - n_\mathrm{surf} -
n_\mathrm{SAM}$ for CF$_3$S on Ag(111), (a) as function of $z$, averaged over the $xy$ plane, in
units of $\rm{\AA}^{-3}$; (b) as an isodensity surface; (c), (d) the same for CH$_3$S on Ag(111).
\label{fig:difcharge}}
\end{figure}

In order to visualize the charge reordering at the surface upon adsorption of the SAM, we calculate
the difference electron density $\Delta n$. It is obtained by subtracting from the total electron
density $n_\mathrm{tot}$ of the SAM on the surface, the electron density $n_\mathrm{surf}$ of the
clean surface and that of the free standing SAM $n_\mathrm{SAM}$. $n_\mathrm{surf}$ and
$n_\mathrm{SAM}$ are obtained in two separate calculations of a clean surface and a free standing
SAM, respectively, with their structures frozen in the adsorbed geometry. As an example,
Fig.~\ref{fig:difcharge} shows $\Delta n$ for SAMs of CF$_3$S and CH$_3$S on Ag(111).

\begin{table*}[!]
\caption{Dipole per molecule $\Delta\mu$, from the change in work function upon adsorption. The
(perpendicular) molecular dipole moment $\mu_\mathrm{SAM}$ in a free standing SAM. The
chemisorption dipole moment is $\mu_\mathrm{chem}=\Delta\mu-\mu_\mathrm{SAM}$. All values are in
D.\label{table:dip}}
\begin{ruledtabular}
\begin{tabular}{lcccccccccccc}
&\multicolumn{4}{c}{Ag} &\multicolumn{4}{c}{Au} &\multicolumn{4}{c}{Pt} \\
& CH$_3$S & C$_2$H$_5$S & CF$_3$S & CF$_3$CH$_2$S & CH$_3$S & C$_2$H$_5$S & CF$_3$S & CF$_3$CH$_2$S
& CH$_3$S & C$_2$H$_5$S & CF$_3$S & CF$_3$CH$_2$S\\ \hline
$\Delta\mu        $   & $-0.32$ & $-0.22$ & $0.97$  & $1.07$ & $-0.86$ & $-0.79$ & $0.43$  & $0.61$ & $-1.28$ & $-1.27$ &$-0.08$  & $0.02$\\
$\mu_\mathrm{SAM} $   & $-0.88$ & $-0.79$ & $0.44$  & $0.50$ & $-0.88$ & $-0.81$ & $0.44$  & $0.53$ & $-0.86$ & $-0.80$ & $0.37$  & $0.47$\\
$\mu_\mathrm{chem}$   &  $0.56$ &  $0.57$ & $0.53$  & $0.57$ &  $0.02$ &  $0.02$ &$-0.01$  & $0.08$ & $-0.42$ & $-0.47$ &$-0.45$  &$-0.45$\\
\end{tabular}
\end{ruledtabular}
\end{table*}

Fig.~\ref{fig:difcharge} illustrates that $\Delta n$ is localized mainly at the metal-SAM
interface, i.e. near the sulphur atoms and the metal atoms in the first surface layers. In case of
adsorption on Ag, electrons are transferred from the metal to the molecule, which results in an
increase of the electron density on the sulphur atoms and a decrease on the surface metal atoms.
The charge transfer does not depend strongly on the molecule, compare
Figs.~\ref{fig:difcharge}(a,b) to (c,d). This is somewhat surprising. since the electronegativity
of CF$_3$S is much higher than that of CH$_3$S.

Very often a charge transfer between two systems is interpreted in terms of their relative
electronegativity. For a metal surface the latter is simply the work function $W_\mathrm{clean}$.
For a molecule the Mulliken electronegativity $\chi_M$ is defined as the average of the ionization
potential and the electron affinity and considered to be the molecular equivalent of a chemical
potential \cite{Parr:89}. We find $\chi_M = 5.4$ eV for the CH$_3$S and CH$_3$CH$_2$S molecules.
Since $\chi_M$ is close to $W_\mathrm{clean}$ for Au(111), this would explain the lack of electron
transfer upon adsorption of these molecules \cite{Derenzi:prl05,Konopka:prl05}. However, the
calculated $\chi_M$ for CF$_3$S and CF$_3$CH$_2$S are much higher, i.e. $6.9$ eV and $6.1$ eV,
respectively. Yet this does not result in a markedly increased electron transfer to these
molecules, as Fig.~\ref{fig:difcharge} indicates. It means that $\chi_M$ is not a generally
suitable parameter to predict the amount of charge transfer between surface and molecules. $\chi_M$
reflects the relative stability of charged molecular states. In particular, for the thiolates
$\chi_M$ reflects the ability of the (fluorinated) alkyl chains to stabilize or screen charge that
resides on the sulphur atom. We suggest that this is not important in case of adsorbed molecules,
as the metal surface takes over this role.

Meanwhile, Fig.~\ref{fig:difcharge} suggests the following analysis. From the change in the work
function upon adsorption of the SAM, $\Delta W = W-W_\mathrm{clean}$, see Table~\ref{table:wf}, one
can obtain the change of the surface dipole upon adsorption, $\Delta \mu = \varepsilon_0 A \Delta
W/e$ (with $\varepsilon_0$ the permittivity of vacuum and $A$ the area of the surface unit cell).
Since the unit cell contains one molecule, $\Delta \mu$ is the change in the surface dipole per
adsorbed molecule. The results are shown in Table~\ref{table:dip}. $\Delta \mu$ contains
contributions from the charge reordering at the interface due to chemisorption, as well as from the
dipole moments of the individual molecules.

The latter can be accounted for by calculating the dipole moment $\mu_\mathrm{SAM}$ per molecule of
free standing SAMs, i.e. without the presence of a metal surface. We focus upon the component of
the dipole that is perpendicular to the surface, since the other components do not contribute to
the work function. As the calculation uses a full monolayer of molecules, it incorporates the
effect on each molecule of the depolarizing field caused by the dipoles of all surrounding
molecules. The calculated $\mu_\mathrm{SAM}$ are given in Table~\ref{table:dip}. The structure of a
SAM is fixed in its adsorption geometry, which is similar for the three metal surfaces. Therefore,
the $\mu_\mathrm{SAM}$ values for adsorption on Ag, Au, and Pt in Table~\ref{table:dip} differ only
slightly. Of course $\mu_\mathrm{SAM}$ depends upon the molecule. In CH$_3$S and CH$_3$CH$_2$S the
dipole points from the sulphur atom to the alkyl group. The large electronegativity of fluor causes
a reversal of the dipole in CF$_3$S and CF$_3$CH$_2$S.

We define the contribution to the interface dipole resulting from chemisorption as
$\mu_\mathrm{chem} = \Delta \mu - \mu_\mathrm{SAM}$. The results shown in Table~\ref{table:dip}
clearly demonstrate that $\mu_\mathrm{chem}$ is nearly independent of the molecule and strongly
dependent on the metal substrate. As an independent check we have also calculated the dipole on the
basis of the electron density redistribution, see Fig.~\ref{fig:difcharge}, $\mu_{\Delta n} =
-e\int \!\!\! \int \!\!\! \int_\mathrm{cell} z \Delta n (\mathbf{r}) dx dy dz$. We find that
$\mu_{\Delta n} \approx \mu_\mathrm{chem}$, which indicates the consistency of this analysis.

The results obtained allow for a simple qualitative picture. The chemisorption dipole
$\mu_\mathrm{chem}$ is very small for all SAMs on Au(111), indicating that the charge transfer
between the Au surfaces and the molecules is small. This generalizes previous results obtained for
methyl thiolate SAMs on Au(111) \cite{Derenzi:prl05,Konopka:prl05}. Since the work function of
Ag(111) is substantially lower than that of Au(111), a significant electron transfer takes place
from the surface to the molecules for SAMs on Ag. This is confirmed by the values of
$\mu_\mathrm{chem}$ for Ag in Table~\ref{table:dip}. Fig.~\ref{fig:difcharge} shows that the
electrons are transferred mainly to the sulphur atoms. Integrating the positive peak of $\Delta n$
on the sulphur atom gives a charge of $(-0.24\pm 0.02)e$. The sign of the charge transfer is such
that $\mu_\mathrm{chem}$ increases the work function with respect to clean Ag(111). By a similar
argument, since the work function of Pt(111) is much higher than that of Au(111), an electron
transfer takes place from the molecules to the surface for adsorption on Pt. The values of
$\mu_\mathrm{chem}$ for Pt in Table~\ref{table:dip} confirm this. In this case the net charge on
the sulphur atom is positive and $\mu_\mathrm{chem}$ decreases the work function with respect to
clean Pt(111).

\begin{figure}[!tbp]
\includegraphics[width=8.5cm]{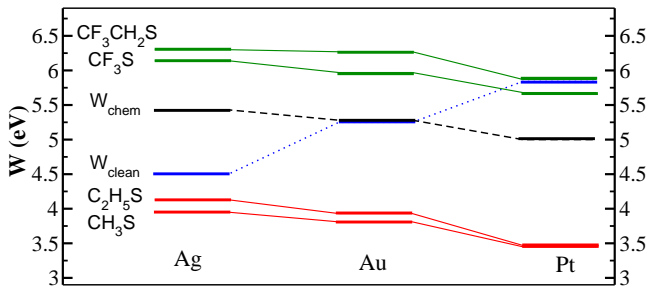}
\caption{ (Color online) Work functions of clean surfaces $W_\mathrm{clean}$ (blue), including the
chemisorption dipole $W_\mathrm{chem}$ (black), and of the SAM covered surfaces (red, green).
\label{fig:wfchem}}
\end{figure}

The size of the charge transfer is remarkable. Chemisorption creates an interface dipole
$\mu_\mathrm{chem}$ that overcompensates for the difference between the metal work functions. We
define a work function that includes the contribution from the chemisorption dipoles as
$W_\mathrm{chem}=W_\mathrm{clean}+e\mu_\mathrm{chem}/(\varepsilon_0 A)$. The results shown in
Fig.~\ref{fig:wfchem} demonstrate that $W_\mathrm{chem}$ decreases in the order Ag, Au and Pt,
whereas $W_\mathrm{clean}$ increases in that order. The work function of the SAM covered surfaces
can then be expressed as $W=W_\mathrm{chem}+e\mu_\mathrm{SAM}/(\varepsilon_0 A)$. From the polarity
of the molecules discussed above, it is clear that SAMs of CH$_3$S and CH$_3$CH$_2$S decrease the
work function, whereas SAMs of CF$_3$S and CF$_3$CH$_2$S increase it.

We thank G. Giovannetti for the molecular calculations and B. de Boer, P. W. M. Blom and P. J.
Kelly for very helpful discussions. This work is part of the research program of the ``Stichting
voor Fundamenteel Onderzoek der Materie" (FOM) and the use of superconmputer facilities was
sponsored by the "Stichting Nationale Computer Faciliteiten" (NCF), both financially supported by
the ``Nederlandse Organisatie voor Wetenschappelijk Onderzoek" (NWO).


\begin{thebibliography}{29}
\bibitem[{fn:()}]{fn:SAMexp}
\bibinfo{note}{This could be due to a number of reasons. The experiments are
  not performed in UHV, which might introduce impurities. Moreover, since in
  experiment longer molecules with a larger polarizability are used, the effect
  of the depolarizing field of the molecular dipoles is larger, which reduces
  the interface dipole.}
\end{thebibliography}
\end{document}